\documentclass[aps,prl,twocolumn,superscriptaddress,showpacs,preprintnumbers,amsmath,amssymb]{revtex4}

\usepackage{graphicx} 
\usepackage{dcolumn}  
\usepackage{color}    
\usepackage{bm}       
\usepackage{ulem}     

\graphicspath{{ps}}

\def\BBbar{B\bar{B}}
\def\qqbar{q\bar{q}}

\def\pim{\pi^-}

\def\kpp{K^{+}\pim\pim}

\def\epem{e^+ e^-}
\def\Kstz{K^{*0}}

\def\lplp{\ell^+\ell^{\prime+}}
\def\Dll{D^-\lplp}
\def\BtoDll{B^+\to \Dll}
\def\Dtokpipi{D^-\to K^+\pi^-\pi^-}
\def\ee{e^+e^+}
\def\emu{e^+\mu^+}
\def\mm{\mu^+\mu^+}
\def\Dee{D^-\ee}
\def\Dem{D^-\emu}
\def\Dmm{D^-\mm}

\def\Y4S{\Upsilon(4S)}
\def\BF{{\cal B}}

\def\GeV{{\rm GeV}}
\def\GeVc{{\rm GeV\!/}c}
\def\GeVcc{{\rm GeV\!/}c^2}

\def\MeVcc{{\rm MeV\!/}c^2}

\def\dE{\Delta E}
\def\Mbc{M_{\rm bc}}
\def\Mkpipi{M_{K\pi\pi}}
\def\cosb{\cos \theta_B}
\def\cossb{\cos^2 \theta_B}
\def\Emiss{E_{\rm miss}}
\def\ndz{\delta z}

\def\btoc{b\to c}

\def\DL2{\Delta L = 2}
\def\lr{\cal R}
\def\0nbb{0\nu \beta\beta}

\def\ulnu{B \to X_u \ell\nu}

\def\ULDee{2.6\times 10^{-6}}
\def\ULDem{1.8\times 10^{-6}}
\def\ULDmm{1.0\times 10^{-6}}

\renewcommand{\arraystretch}{1.1}
\begin{document}

\preprint{\vbox{ 
    \hbox{   } 
    \vspace{0.5cm}
    \hbox{Belle Draft ver 2.1} 
    \hbox{\today} 
}}

\title{\quad\\[0.5cm] 
Search for Lepton-number-violating $\BtoDll$ Decays}


\affiliation{Budker Institute of Nuclear Physics SB RAS and Novosibirsk State University, Novosibirsk 630090}
\affiliation{Faculty of Mathematics and Physics, Charles University, Prague}
\affiliation{University of Cincinnati, Cincinnati, Ohio 45221}
\affiliation{Justus-Liebig-Universit\"at Gie\ss{}en, Gie\ss{}en}
\affiliation{Gifu University, Gifu}
\affiliation{Gyeongsang National University, Chinju}
\affiliation{Hanyang University, Seoul}
\affiliation{University of Hawaii, Honolulu, Hawaii 96822}
\affiliation{High Energy Accelerator Research Organization (KEK), Tsukuba}
\affiliation{Hiroshima Institute of Technology, Hiroshima}
\affiliation{Indian Institute of Technology Guwahati, Guwahati}
\affiliation{Indian Institute of Technology Madras, Madras}
\affiliation{Institute of High Energy Physics, Chinese Academy of Sciences, Beijing}
\affiliation{Institute of High Energy Physics, Vienna}
\affiliation{Institute of High Energy Physics, Protvino}
\affiliation{Institute for Theoretical and Experimental Physics, Moscow}
\affiliation{J. Stefan Institute, Ljubljana}
\affiliation{Kanagawa University, Yokohama}
\affiliation{Institut f\"ur Experimentelle Kernphysik, Karlsruher Institut f\"ur Technologie, Karlsruhe}
\affiliation{Korea Institute of Science and Technology Information, Daejeon}
\affiliation{Korea University, Seoul}
\affiliation{Kyungpook National University, Taegu}
\affiliation{\'Ecole Polytechnique F\'ed\'erale de Lausanne (EPFL), Lausanne}
\affiliation{Faculty of Mathematics and Physics, University of Ljubljana, Ljubljana}
\affiliation{University of Maribor, Maribor}
\affiliation{Max-Planck-Institut f\"ur Physik, M\"unchen}
\affiliation{University of Melbourne, School of Physics, Victoria 3010}
\affiliation{Nagoya University, Nagoya}
\affiliation{Nara Women's University, Nara}
\affiliation{National Central University, Chung-li}
\affiliation{National United University, Miao Li}
\affiliation{Department of Physics, National Taiwan University, Taipei}
\affiliation{H. Niewodniczanski Institute of Nuclear Physics, Krakow}
\affiliation{Nippon Dental University, Niigata}
\affiliation{Niigata University, Niigata}
\affiliation{University of Nova Gorica, Nova Gorica}
\affiliation{Osaka City University, Osaka}
\affiliation{Pacific Northwest National Laboratory, Richland, Washington 99352}
\affiliation{Panjab University, Chandigarh}
\affiliation{Research Center for Nuclear Physics, Osaka}
\affiliation{Saga University, Saga}
\affiliation{University of Science and Technology of China, Hefei}
\affiliation{Seoul National University, Seoul}
\affiliation{Sungkyunkwan University, Suwon}
\affiliation{School of Physics, University of Sydney, NSW 2006}
\affiliation{Tata Institute of Fundamental Research, Mumbai}
\affiliation{Excellence Cluster Universe, Technische Universit\"at M\"unchen, Garching}
\affiliation{Toho University, Funabashi}
\affiliation{Tohoku Gakuin University, Tagajo}
\affiliation{Tohoku University, Sendai}
\affiliation{Department of Physics, University of Tokyo, Tokyo}
\affiliation{Tokyo Institute of Technology, Tokyo}
\affiliation{Tokyo Metropolitan University, Tokyo}
\affiliation{Tokyo University of Agriculture and Technology, Tokyo}
\affiliation{CNP, Virginia Polytechnic Institute and State University, Blacksburg, Virginia 24061}
\affiliation{Yonsei University, Seoul}
  \author{O.~Seon}\affiliation{Nagoya University, Nagoya} 
  \author{Y.-J.~Kwon}\affiliation{Yonsei University, Seoul} 
  \author{T.~Iijima}\affiliation{Nagoya University, Nagoya} 

  \author{I.~Adachi}\affiliation{High Energy Accelerator Research Organization (KEK), Tsukuba} 
  \author{H.~Aihara}\affiliation{Department of Physics, University of Tokyo, Tokyo} 
  \author{D.~M.~Asner}\affiliation{Pacific Northwest National Laboratory, Richland, Washington 99352} 
  \author{T.~Aushev}\affiliation{Institute for Theoretical and Experimental Physics, Moscow} 
  \author{A.~M.~Bakich}\affiliation{School of Physics, University of Sydney, NSW 2006} 
  \author{E.~Barberio}\affiliation{University of Melbourne, School of Physics, Victoria 3010} 
  \author{A.~Bay}\affiliation{\'Ecole Polytechnique F\'ed\'erale de Lausanne (EPFL), Lausanne} 
% ...
\collaboration{The Belle Collaboration}


\begin{abstract}
  We perform the first search for lepton-number-violating 
  $\BtoDll$ decays, where $\ell$ and $\ell^\prime$ stand for $e$ or $\mu$,
  using 772$\times 10^6$ $\BBbar$ pairs accumulated at the $\Upsilon(4S)$
  resonance with the Belle detector at the KEKB $e^+e^-$ collider.  No
  evidence for these decays has been found. Assuming uniform
  three-body phase space distributions for the $\Dll$ decays, 
  we set the following upper limits on the branching fractions at 90\% 
  confidence level:
  ${\cal B}(B^+ \to \Dee) < \ULDee$, ${\cal B}(B^+ \to \Dem) <
  \ULDem$ and ${\cal B}(B^+ \to \Dmm) < \ULDmm$.
\end{abstract}



{\maketitle
\begin{figure}[hbt]
  (a)\rule{6.5cm}{0cm}\\[-3mm]
  \includegraphics[width=0.33\textwidth]{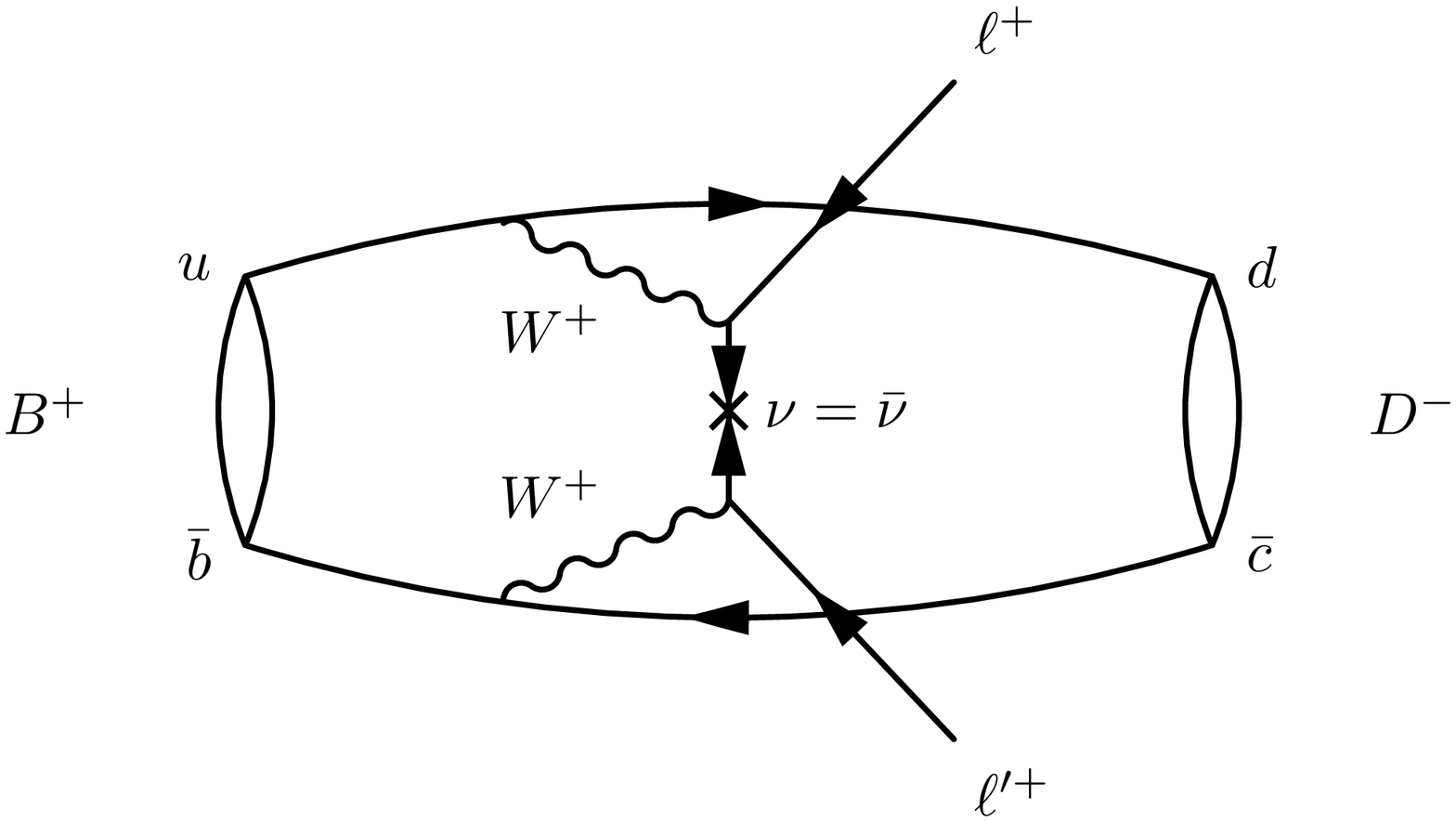}\\[2mm]
  (b)\rule{6.5cm}{0cm}\\[-3mm]
  \includegraphics[width=0.33\textwidth]{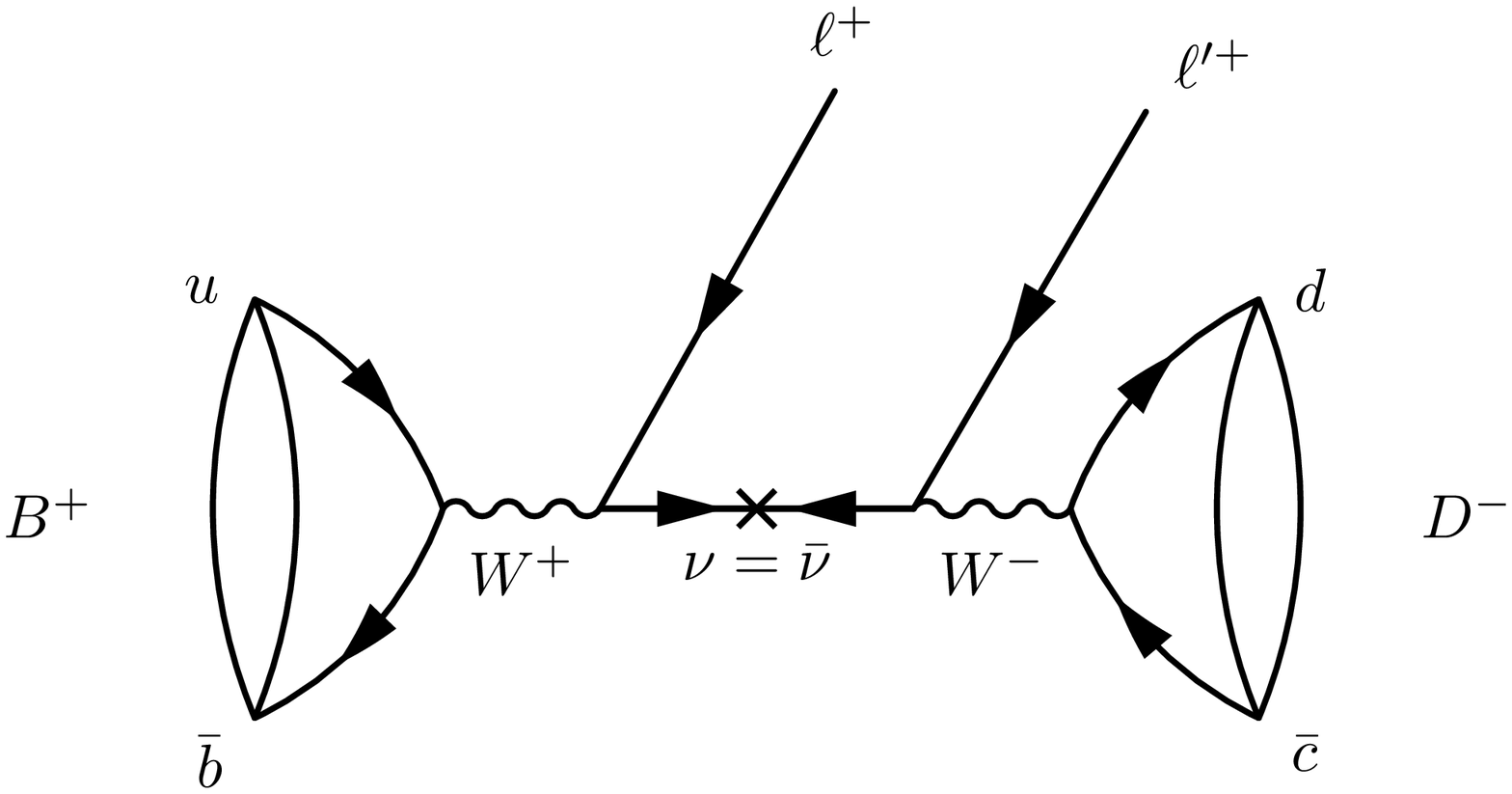}
  \caption{Feynman diagrams for $\BtoDll$.}
  \label{Feyn}
\end{figure}


{\renewcommand{\thefootnote}{\fnsymbol{footnote}}}
\setcounter{footnote}{0}  

In the Standard Model (SM) neutrinos are left-handed massless
particles and lepton number is conserved. 
However, the strong evidence for neutrino oscillations~\cite{mixing}
indicates that neutrinos do have non-zero
masses.  An important question then arises regarding 
the origin of neutrino masses: whether they are of
Dirac or Majorana type. If neutrinos are purely of Dirac type, they
must have right-handed singlet components in addition to the
left-handed states required in order to accommodate neutrino masses. In
this case, lepton number is conserved. On the other hand, if
there are Majorana-type neutrino states, a neutrino cannot be
distinguished from its own antiparticle. As a result,
lepton-number-violating processes can occur in which lepton
number changes by two units ($\DL2$).  

There have been many experimental attempts to search for $\DL2$
processes. The most thoroughly tested of these processes are
neutrinoless nuclear double beta decays ($\0nbb$)~\cite{0nbb}. While
the experiments are very sensitive, uncertainties in the nuclear
matrix elements for $\0nbb$ would make it difficult to extract the 
mass scale of the neutrinos involved in such decays.
As an alternative, several authors have considered $\DL2$ processes in 
meson decays~\cite{AtreHan,ZhangWang,Cvetic}. 

The only existing experimental result for $\DL2$ 
$B$ meson decays is that of the CLEO collaboration,
which  searched for 
$B^+ \to h^-\lplp$~\cite{chargeconj},
where $h$ stands for $\pi$, $K$, $\rho$, 
or $K^*$ and $\ell$ stands for $e$ or $\mu$. They set
upper limits on branching fractions for these decays in
the range of $(1.0-8.3)\times 10^{-6}$ at 
90\% confidence level (CL)~\cite{CLEOhll}. 
Since $b\to c$ decays are in general favored in comparison to 
charmless $B$ decays, it is interesting to extend the search 
to $\BtoDll$ decays. 
Two well-known diagrams for such decays are shown in Fig. \ref{Feyn} (a) and (b). 
According to theoretical calculations, with a heavy Majorana neutrino of mass 
within the $(2-4)~\GeVcc$ range, the branching fractions of $\BtoDll$ can be
larger than $10^{-7}$~\cite{ZhangWang,Cvetic} with the diagram 
in Fig. \ref{Feyn} (b) giving the dominant contribution.

In this paper, we report the first searches for the $B^+ \to \Dee$,
$\Dem$ and $\Dmm$ decays. 
The results are based on a data sample containing $772 \times
10^6$ $\BBbar$ pairs collected at the $\Upsilon(4S)$ resonance
with the Belle detector at the KEKB~\cite{KEKB} asymmetric-energy $\epem$ collider 
(3.5 on 8 GeV).
The Belle detector is a large-solid-angle magnetic spectrometer
consisting of a silicon vertex detector, a 50-layer central
drift chamber (CDC), an array of aerogel threshold Cherenkov counters
(ACC), a time-of-flight scintillation counter (TOF), and an array of
CsI(Tl) crystals for an electromagnetic calorimeter (ECL) located inside
a superconducting solenoid coil that provides a 1.5 T magnetic field.
An iron flux-return located outside the solenoid is equipped with
resistive plate chambers to identify muons as well as $K_L^0$ mesons
(KLM). The Belle detector is described in detail elsewhere~\cite{Belle}.

The analysis procedure is established using Monte Carlo (MC)
simulations \cite{MC}, as well as data control samples wherever possible. 
Since we have no prior knowledge nor widely-accepted model for the 
decay dynamics of $\BtoDll$, the signal MC samples are generated 
uniformly over the three-body phase space,
and we restrict our analysis and interpretation to this model only.

To reconstruct $\BtoDll$ decays, we first look for an energetic 
same-sign dilepton and combine it with a $D$ candidate 
requiring a proper charge combination for the dilepton. 
All charged tracks are required to originate near the
interaction point and have impact parameters
within 5 cm along the beam direction 
and within 1 cm in the transverse plane to the beam direction. 

Electrons are identified using the energy and shower profile 
in the ECL, the light yield in the ACC ($N_{\rm p.e.}$) 
and the specific ionization energy loss in the CDC (d$E$/d$x$). 
This information is used to form an electron (${\cal L}_e$) 
and non-electron (${\cal L}_{\overline{e}}$) likelihood.
The likelihoods are utilized in the form of a likelihood ratio 
${\lr}_{e} = {\cal L}_{e} / ( {\cal L}_e + {\cal
  L}_{\overline{e}})$ \cite{eid}. 
Applying a requirement on ${\lr}_e$, we select electrons with an efficiency 
and a misidentification rate of approximately 90\% and 0.1\%, respectively, 
in the kinematic region of interest. 
Muons are distinguished from other charged tracks by their ranges 
and their hit profiles in the KLM. This information is utilized 
in a likelihood ratio approach~\cite{muid} similar to the one used 
for the electron identification (ID). We select muons with an efficiency 
and a misidentification rate of approximately 90\% and 1\%, respectively, 
in the kinematic region of interest. 
The efficiencies for electron (muon) ID are evaluated 
from data using the $e^+e^-(\mu^+\mu^-)$ pair 
production via the two-photon reaction $\gamma\gamma\to e^+e^-(\mu^+\mu^-)$.
Since the lepton ID performance is worse for lower-momentum tracks, 
we require the lepton momentum in the laboratory frame to be greater 
than 0.5 $\GeVc$ and 0.8 $\GeVc$ for electrons and muons,
respectively. 

We require a same-sign lepton pair that has a total energy
in the $\Upsilon(4S)$ center-of-mass (CM) frame greater than 1.3 GeV. 
More than 95\% of events have only one same-sign lepton pair.
When there is more than one same-sign lepton pair, 
we choose the most energetic same-sign lepton pair from the three most
energetic leptons in the event. 

Candidate $D^-$ mesons are reconstructed in the $D^-\to \kpp$ decay. 
Kaons and pions are selected from charged particles by applying hadron
ID~\cite{kid}. 
The hadron ID utilizes the time of flight measured in
the TOF as well as $N_{\rm p.e.}$ and d$E$/d$x$ in a likelihood ratio
approach, which is similar to that used for lepton ID.  
We discriminate kaons (pions) from pions (kaons) with 
an efficiency of approximately 91\% (95\%) and a
misidentification rate below 4\% (6\%) in the kinematic region of interest.
The rates are evaluated from data using kinematically
reconstructed $D^{*+}\to D^0\pi^+\to K^-\pi^+\pi^+$ decays.
The three tracks from the $D^-$ candidate are fit
to a common vertex and are required to have a $K^+\pi^-\pi^-$
invariant mass ($\Mkpipi$) within approximately $\pm 10 ~\MeVcc$ 
from the nominal $D^-$ mass~\cite{pdg}. 
The $\Mkpipi$ distribution is fit to two Gaussian functions with 
a common mean. The $\Mkpipi$ mass window is chosen to be 
$\pm 3$ times the width of the narrower Gaussian component.
The average multiplicity of $D^-$ candidates is 1.3 per event.
If there are multiple $D^-$ candidates, 
we choose the one with $\Mkpipi$ closest to the nominal $D$ mass.

The same-sign dilepton and the $D^-$ candidates are 
combined to form a $B$ candidate, and are fit to a common vertex. 
The $B$ candidates are kinematically identified using two variables:
the energy difference, $\dE\equiv E_B - E_{\rm beam}$,
and the beam-energy-constrained $B$ meson mass, 
$\Mbc \equiv \sqrt{E_{\rm beam}^2 - p_{B}^2}$. 
Here, $E_{\rm beam}$ is the beam energy and  $E_B$ 
and $p_{B}$ are the energy and momentum, respectively, 
of a $B$ candidate; these variables are defined in the CM frame.  
We select events with $\Mbc > 5.2 ~\GeVcc$ and $|\dE| < 0.3 ~\GeV$
(``analysis region''). The signal region is defined as 5.27 $\GeVcc 
< \Mbc < 5.29 ~\GeVcc$ and $-0.055~(-0.035)~\GeV < \dE < 0.035~\GeV$ 
for the $\ee$ and $\emu $ modes ($\mm$ mode), respectively. 
For background studies, 
we use a subset of the analysis region that excludes 
the signal region (``background region'').

One of the major backgrounds comes from the 
continuum production of quark pairs $e^+e^-\to\qqbar$ ($q = u,d,s$ and $c$). 
The continuum background is discriminated from the signal by utilizing 
the difference of the event shapes in the CM frame. 
Since $B$ mesons are produced from the $\Upsilon(4S)$ resonance
nearly at rest in the CM frame 
their final state particles are distributed isotropically.
In the continuum, on the other hand, $\qqbar$ pairs 
hadronize back-to-back and give rise to a two-jet-like shape. 
To quantify the event shape characteristics, 
we use Fox-Wolfram moments~\cite{FoxWolfram} with 
modifications optimized for exclusive $B$ decays~\cite{SFW}.
A single discrimination variable, $\cal F$, is obtained by 
applying a linear Fisher discriminant~\cite{Fisher}
to the moments and maximizing their discrimination power.

In addition to $\cal F$, we also use the cosine of the polar angle of the 
$B$ candidate flight direction evaluated in the CM frame ($\cosb$).
Since the $\Upsilon(4S)$ is a vector particle that decays to
a pair of spinless $B$ mesons, the $\cosb$ distribution of the $B$ 
mesons follows a $|Y_{11}|^2 \propto  1-\cossb$ distribution, 
while random track combinations in the  
continuum have a nearly uniform distribution.

The other major background comes from semileptonic
$B$ decays such as $B \to D^- \ell^+ \nu_\ell X$ with $D^-\to \kpp$, 
where $X$ denotes any particle. Such decays can be misreconstructed 
as signal by combining a same-sign
lepton from the decay products of the other $B$.
In such background events, each lepton is produced along with a neutrino, 
resulting in large missing energy, while
the signal tends to have small missing energy because
there are no neutrinos in the final state. 
Here the missing energy, $\Emiss$, is defined as  
$\Emiss \equiv 2E_{\rm beam} - \sum E_{\rm det}$, 
where $\sum E_{\rm det}$ denotes the sum of energies of 
all the detected particles in the event.
Moreover, the same-sign leptons in such background 
events originate from different $B$ mesons. 
As a result, the difference between the impact parameters 
of the two leptons in the beam direction, $\ndz$, 
tends to be larger in such background events than in the signal.
Therefore, we use $\Emiss$ and $\ndz$ as variables to suppress these 
backgrounds.

The four variables, ${\cal F}$, $\cosb$, $\Emiss$ and $\ndz$,
are combined together into a single likelihood ratio 
$\lr_{\rm s} = {\cal L}_{\rm s}/({\cal L}_{\rm s}+{\cal L}_{\rm b})$,
where ${\cal L}_{\rm s(b)}$ denotes the signal (background) likelihood 
defined as the product of the signal (background) probability densities for each of the four variables.
The two major backgrounds can be suppressed 
by applying a requirement on $\lr_{\rm s}$. 
The probability density functions (PDFs) are taken from 
the distributions in the MC samples. The background sample includes 
continuum and $\BBbar$ components, where $B$ decays are 
limited to $\btoc$ decays. The optimal requirement on $\lr_{\rm s}$ is 
determined by maximizing the figure of merit, 
$\epsilon_{\rm s}/\sqrt{N_{\rm b}}$, 
where $\epsilon_{\rm s}$ is the signal efficiency estimated with the signal MC sample, 
and $N_{\rm b}$ is the number of expected background events in the signal region.}
Since only a small number of events remain in the signal region 
after the $\lr_{\rm  s}$ requirement, the value of $N_{\rm b}$ is obtained 
by scaling the number of events in the analysis region using the background MC sample,  
where the scale factor is determined from the same MC sample 
but without the $\lr_{\rm s}$ requirement.
The optimal requirements on $\lr_{\rm s}$ eliminate more than 99\%
of the background while retaining 11-26\%
of the signal depending on the mode.

In addition to the two dominant backgrounds described above, 
we checked backgrounds that might produce a signal-like enhancement in the 
$\Mbc$-$\dE$  distribution having more than one particle misidentified. 
Possible peaking backgrounds include
$B^+\to J/\psi (\to\ell^+\ell^-)K^+\pi^+\pi^-$, 
with the $\ell^-$ and $\pi^+$ misidentified as a $\pi^-$ and $\ell^+$, 
respectively. 
Contributions from these decays are investigated using the MC sample 
that is approximately equivalent to 50 times the luminosity of the data sample. 
The contribution of $B^+ \to D^-h^+h^{\prime +}$ 
decays with both same-sign hadrons ($h^{(\prime)}$) misidentified as leptons is estimated 
from the number of $B^+\to D^-h^+h^{\prime +}$ events weighted by the $h^\prime$
misidentification rates, both evaluated in data.
Background events from misreconstructed $D^-$ mesons are 
studied using the $D^-$ mass sideband.
We studied charmless hadronic $B$ meson decays as well as 
semileptonic $\ulnu$ decays using dedicated high-statistics MC samples, 
which are approximately equivalent to 21 and 14 times the luminosity 
of the data sample, respectively.

After applying the $\lr_{\rm s}$ requirements, 
5, 23 and 40 events remain in the background region for the $\ee$, 
$\emu$ and $\mm$ modes, respectively. 
The background levels are in good agreement 
with the expectations from the background MC samples; 
4, 22 and 38 events, respectively.
The signal region of the data sample is not examined until all the 
selection criteria are fixed and the systematic uncertainties are evaluated.
From the MC samples the signal efficiencies are evaluated
to be 1.2\% - 1.9\%, depending on the mode.
Here the small difference between the MC and data samples on the 
particle ID performance is corrected.
In each case, the correction is approximately 2\% or smaller. 
The expected numbers of background events in the signal region ($N_{\rm exp}^{\rm bkg}$)
are 0.18, 0.83 and 1.44 events 
for the $\ee$, $\emu$ and $\mm$ modes, respectively. 
These background expectations are obtained by scaling the results 
of a two-dimensional fit to the background region,
where we use a common background shape for the three signal modes
to compensate for the low statistics.
The PDFs to fit the background distribution are an
ARGUS function~\cite{argus} for $\Mbc$ and a linear function for $\dE$.
We take the ratio of the integral of the PDF in the signal region 
to that in the background region;
its value and error are 0.036 and 0.006, respectively.

\begin{figure}[hbt]
  \includegraphics[width=0.3\textwidth]{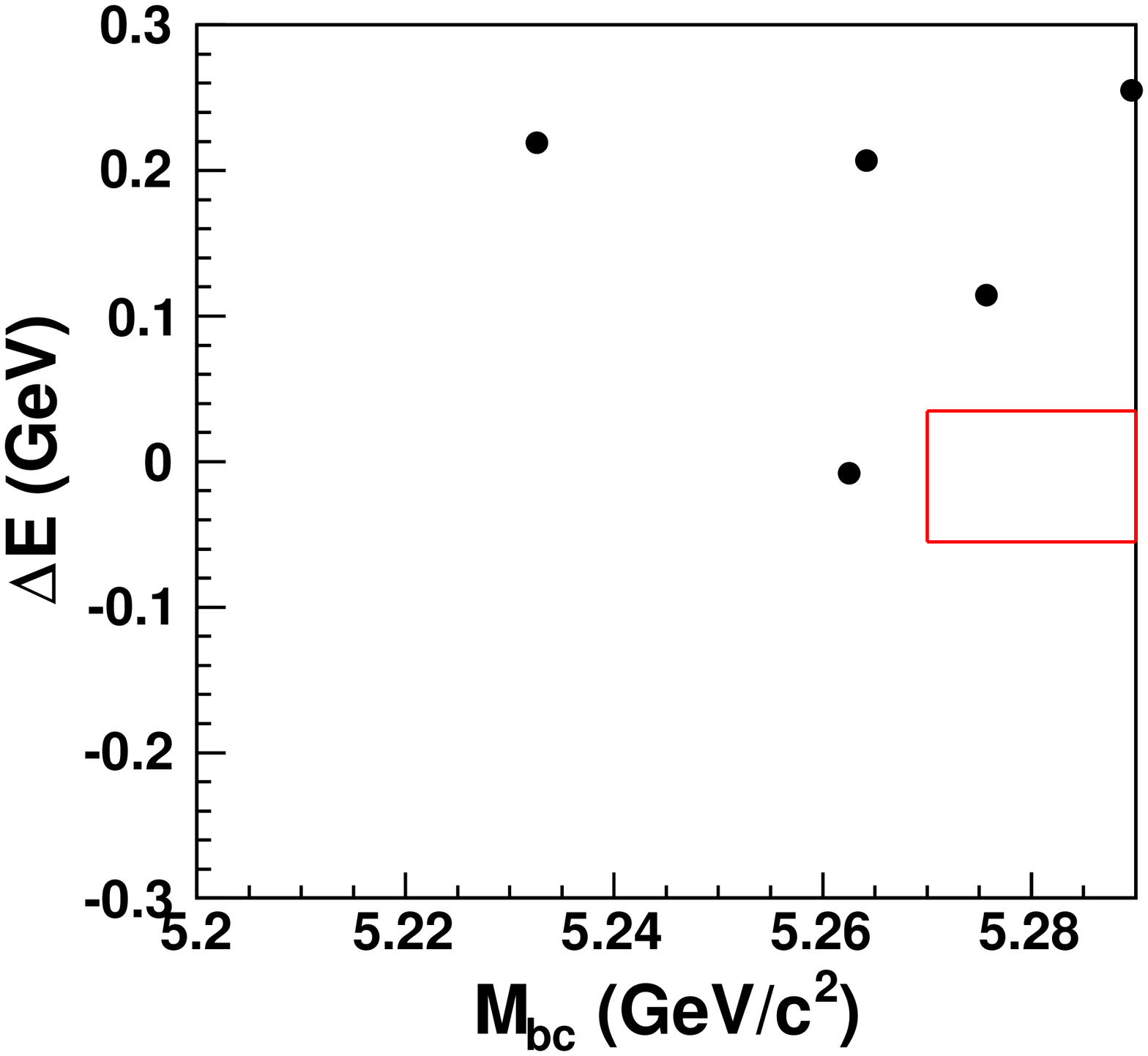}\\
  \includegraphics[width=0.3\textwidth]{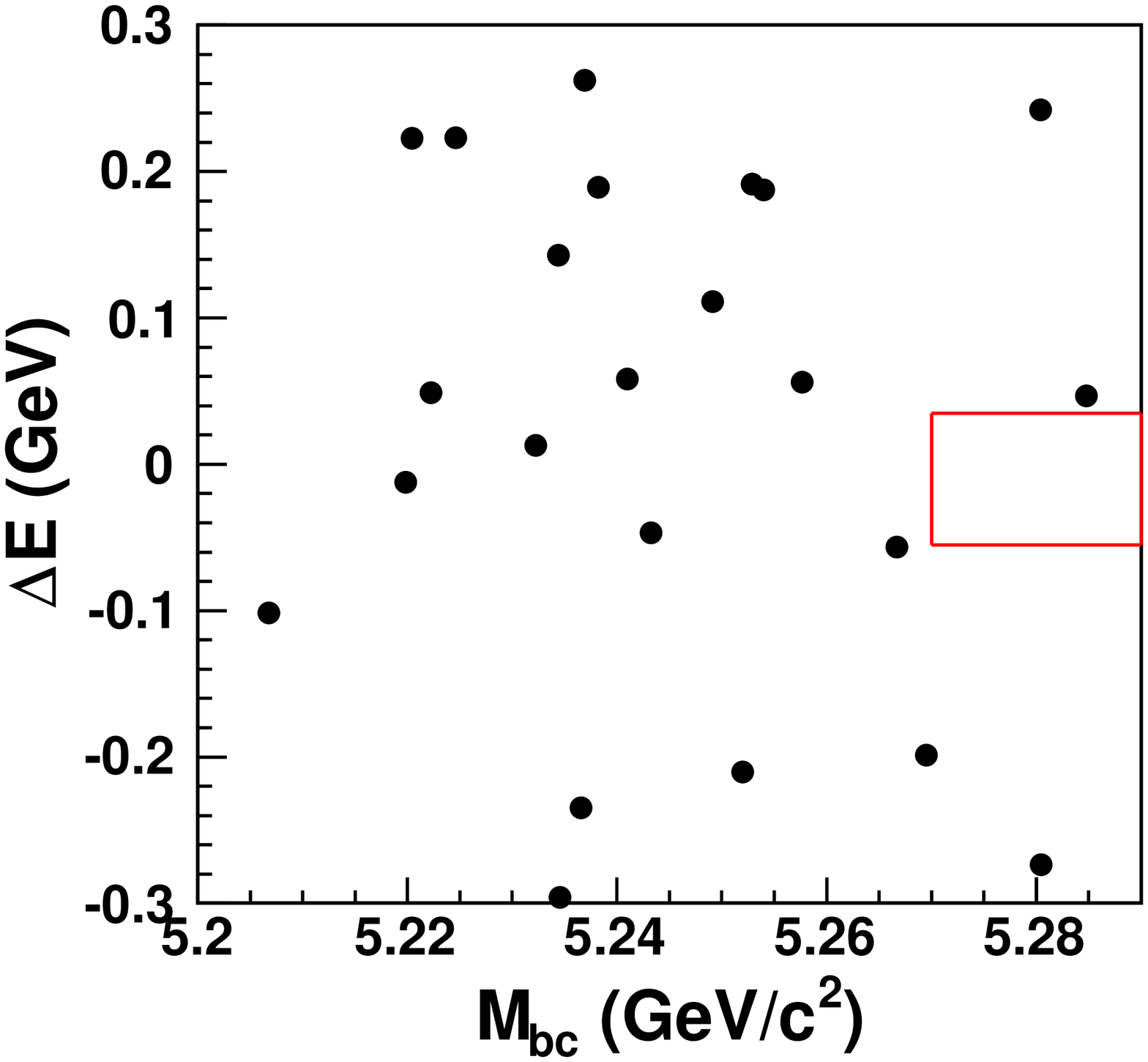}\\
  \includegraphics[width=0.3\textwidth]{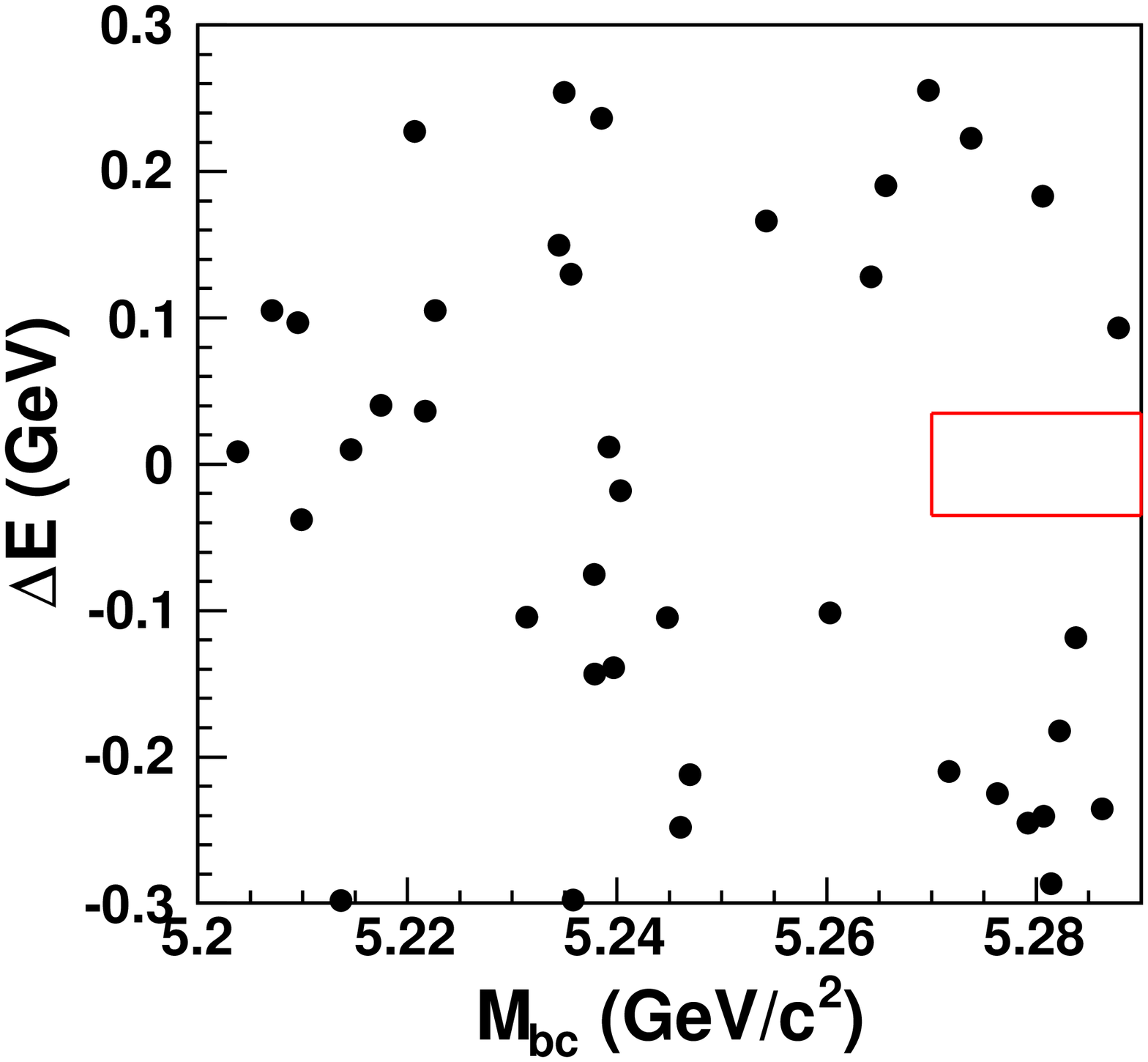}\\
  \caption{The $\Mbc$-$\dE$ distributions of $\Dee$(top), 
    $\Dem$ (middle) and $\Dmm$ (bottom) final states in data.
    The boxes indicate the signal regions.
  }
  \label{figResults}
\end{figure}

Figure~\ref{figResults} shows the $\Mbc$-$\dE$ distributions of 
events in the analysis region of the data sample,
which pass all the selection criteria.
The signal region is unblinded and no events are observed in any mode, 
which is consistent with the background expectations.
Table~\ref{result} summarizes the signal efficiency, 
the number of observed events and the expected number 
of background events in the signal region for each mode. 

The systematic uncertainties on $N_{\rm exp}^{\rm bkg}$ 
are also listed in Table \ref{result}. 
Each of the uncertainties combines the errors 
on the number of events in the background region and on the scale factor. 
For the latter each PDF shape parameter is varied 
by its fit error, 
and the resulting changes of the scale factor are added in quadrature.
The fit procedure and the uncertainty evaluation are 
also applied to the background MC sample. 
Moreover, a mode-dependent PDF shape, 
taken from the background MC sample of each mode, 
is examined in the same manner. As a conservative evaluation, 
the uncertainties obtained with two MC-based PDFs are added in quadrature
in the uncertainty for each mode listed in Table \ref{result}.

Systematic uncertainties for efficiency determination are 
summarized in Table~\ref{system}. 
They are dominated by the tracking efficiency and the 
requirement of $\lr_{\rm s}$.
The uncertainty on the tracking efficiency is 
obtained by comparing partially and fully reconstructed
$D^{*+}\to\pi^+D^0$, $D^0\to K^0_{S}(\to \pi^+\pi^-)\pi^+\pi^-$
decays in data and MC simulation.
The systematic uncertainties on the particle ID 
efficiencies are evaluated using the data control samples mentioned earlier.
The uncertainty on the selection efficiency of the $\lr_{\rm s}$ 
requirements is evaluated from the ratio of the number of events 
in the signal region before and after applying the $\lr_{\rm s}$ requirement for data 
and MC samples using the $B^0\to J/\psi \Kstz$ mode.
The number of events in the control sample is extracted by applying 
the 2-dimensional fit described earlier with a PDF component for the corresponding decay.
Since this control sample does not represent the $\emu$ mode very well, 
we take the larger of the two dilepton mode uncertainties for the $\emu$ mode.
The same control sample is used to evaluate the uncertainty 
on the efficiency of the signal region acceptance.
The same evaluation is applied for the uncertainty on the efficiency of the
$\Mkpipi$ acceptance.
A difference between the $\Mkpipi$ shapes in data and MC 
would result in the different event fractions in the signal region. 
The control sample used is $B^0\to D^-(\to K^+\pi^-\pi^-)\pi^+$, 
which is kinematically reconstructed after applying hadron ID requirements.

No events are observed in the signal region. We set upper limits on
the branching fractions based on a frequentist approach \cite{FeldmanCousins}.
We calculate the 90\% C.L. upper limit on the branching fractions 
including systematic uncertainty, using the POLE program without conditioning \cite{POLE}.
Except for the uncertainty on $N_{\rm exp}^{\rm bkg}$, all the 
systematic uncertainties, 
including those on the number of $\BBbar$ events ($N_{\BBbar}$) and on 
the branching fraction 
of $D^-\to K^+\pi^-\pi^-$~\cite{pdg}, are assigned to multiplicative quantities 
in the upper limit calculation. 
These are found to be 8.8\%, 9.8\% and 9.7\% for the $\ee$, $\emu$ and 
$\mm$ modes, respectively, as summarized in Table \ref{system}. 
The  90\% CL upper limits are  $(1.0 - 2.6)\times 10^{-6}$ 
depending on the mode, as listed in Table \ref{result}.

In summary, we have searched the lepton-number-violating 
$\BtoDll$ decays for the first time. We find no signal candidates.
Assuming uniform three-body phase space distributions,
we set the following upper limits
on the branching fractions at $90\%$ CL: ${\cal B}(B^+ \to \Dee) <
\ULDee$, ${\cal B}(B^+ \to \Dem) < \ULDem$, and ${\cal B}(B^+ \to
\Dmm) < \ULDmm$. 

We thank the KEKB group for excellent operation of the accelerator,
the KEK cryogenics group for efficient solenoid operations, and the
KEK computer group and the  for valuable computing and SINET4
network support. We acknowledge support from MEXT, JSPS and Nagoya's
TLPRC (Japan); ARC and DIISR (Australia); NSFC (China); MSMT (Czechia);
DST (India); MEST, NRF, NSDC of KISTI, and WCU (Korea); MNiSW (Poland); 
MES and RFAAE (Russia); ARRS (Slovenia); SNSF (Switzerland); 
NSC and MOE (Taiwan); and DOE (USA). 
O.~S. acknowledges support by the COE program of Japan. 
Y.-J.~K. acknowledges support by NRF Grant No. 2010-0015967.

\begin{table}[hbt]
  \renewcommand{\arraystretch}{1.3}
  \caption{Results of the $\BtoDll$ search; 
    $\epsilon$ is the signal reconstruction efficiency, $N_{\rm obs}$ is the
    number of events in the signal region, $N_{\rm exp}^{\rm bkg}$ 
    is the expected number of background events in 
    the signal region, and U.L. is the 90\% CL upper
    limit on the branching fraction.
    The efficiencies shown in the table do not include the 
    branching fraction of the $D^-$ decay.
    }
  \label{result} 
  \begin{ruledtabular}
    \begin{tabular}{ccccc}
      Mode & $\epsilon$ [\%] & $N_{\rm obs}$ & $N^{\rm bkg}_{\rm exp}$
      & U.L. [$10^{-6}$]\\\hline 
      $B^+\to\Dee$ & 1.2 & 0 & 0.18$\pm$0.13 & $<$ 2.6 \\
      $B^+\to\Dem$ & 1.3 & 0 & 0.83$\pm$0.29 & $<$ 1.8 \\
      $B^+\to\Dmm$ & 1.9 & 0 & 1.44$\pm$0.43 & $<$ 1.0 \\
    \end{tabular}
  \end{ruledtabular}
\end{table}


\begin{table}[hbt]
\renewcommand{\arraystretch}{1.3}
  \caption{Summary of multiplicative systematic uncertainties. 
    The units are in percent.}
  \label{system}
  \begin{ruledtabular}
    \begin{tabular}{ccccc}
      Source & $\Dee$ & $\Dem$ & $\Dmm$ \\\hline
      MC statistics          & $<$ 0.1 & $<$ 0.1 & $<$ 0.1 \\
      Tracking efficiency    & 5.2 & 5.2 & 5.2 \\ 
      Lepton ID              & 3.1 & 3.5 & 3.6 \\
      Hadron ID          & 1.4 & 1.4 & 1.4\\ 
      $\Mbc$ and $\dE$       & 2.0 & 2.0 & 1.5 \\
      $\Mkpipi$              & 2.4 & 2.5 & 2.4\\ 
      $\lr_{\rm s}$          & 3.0 & 4.9 & 4.9\\
      $N_{\BBbar}$           & 1.4 & 1.4 & 1.4\\
      $\BF(\Dtokpipi)$ & 4.3 & 4.3 & 4.3\\\hline 
      Sum                    & 8.8 & 9.8 & 9.7\\
    \end{tabular}
  \end{ruledtabular}
\end{table}


\begin{thebibliography}{99}

\bibitem{mixing} Y. Fukuda {\it et al.} (Super-Kamiokande Collaboration), Phys. Rev. Lett. {\bf 81}, 1562 (1998); 
  Y. Fukuda {\it et al.} (Super-Kamiokande Collaboration), Phys. Lett. B {\bf 539}, 179 (2002);
  Q. R. Ahmad {\it et al.} (SNO Collaboration), Phys. Rev. Lett. {\bf 89}, 011301 (2002);
  Eguchi K, {\it et all.} (KamLAND Collaboration), Phys. Rev. Lett. {\bf 94} 081801 (2005).

\bibitem{0nbb} For a recent review see, e.g., F. T. Avignone III 
  {\it et al.}, Rev. Mod. Phys. {\bf 80}, 481 (2008) and references therein.
  
\bibitem{AtreHan} A. Atre {\it et al.}, JHEP, {\bf 0905}, 030 (2009).
  
\bibitem{ZhangWang} J.-M. Zhang and G.-L. Wang, arXiv:1003.5570 [hep-ph].
  
\bibitem{Cvetic} G. Cvetic {\it et al.}, Phys. Rev. D {\bf 82},
  053010 (2010).
  
\bibitem{chargeconj} Throughout this paper, charge-conjugate processes
  are implied unless explicitly stated otherwise. 
  
\bibitem{CLEOhll} K. W. Edwards {\it et al.} (CLEO collaboration),
  Phys. Rev. D {\bf 65}, 111102 (2002). 
  
\bibitem{KEKB} 
  S.~Kurokawa and E.~Kikutani, Nucl. Instr. and Meth. A {\bf 499}, 1
  (2003), and other papers in this volume.
  
\bibitem{Belle}
  A.~Abashian {\it et al.} (Belle Collaboration),
  Nucl. Instr. and Meth. A {\bf 479}, 117 (2002).
  
\bibitem{MC}
  We use the EvtGen 
  package to generate MC events,
  D.J. Lange, Nucl. Instr. and Meth. A {\bf 462}, 152 (2001). 
  The detector simulation utilizes the GEANT package,
  R. Brun {\it et al.}, GEANT 3.21, CERN Report DD/EE/84-1 (1984).
  
\bibitem{eid}
    K. Hanagaki {\it et al.}, 
    Nucl. Inst. and Meth. A {\bf 485}, 490 (2002).

\bibitem{muid}
    A. Abashian {\it et al.},
    Nucl. Inst. and Meth. A {\bf 491}, 69 (2002).
  
\bibitem{kid} E. Nakano {\it el al.}, 
    Nucl. Inst. and Meth. A {\bf 494}, 402 (2002).
  
\bibitem{pdg} K. Nakamura {\it et al.} (Particle Data Group), 
  J. Phys. G {\bf 37}, 075021 (2010).

\bibitem{FoxWolfram} 
  G.C. Fox and S. Wolfram, 
  Phys. Rev. Lett. {\bf 41}, 1581 (1978).

\bibitem{SFW}
  S. H. Lee {\it et al.} (Belle Collaboration), 
Phys. Rev. Lett. {\bf 91}, 261801 (2003).

\bibitem{Fisher}
  R. A. Fisher, Ann. Eugen. {\bf 7}, 179 (1936).
  

\bibitem{argus}
  H.~Albrecht {\it et al.} (ARGUS Collaboration),
  Phys. Lett. B {\bf 241}, 278 (1990).
  
\bibitem{FeldmanCousins}
  G.J. Feldman and R.D. Cousins, Phys. Rev. D {\bf 57}, 3873 (1998).

\bibitem{POLE} J. Conrad {\it et al.}, Phys. Rev. D {\bf 67}, 012002
  (2003). 
\end{thebibliography}
\end{document}